\documentclass[aps,prl,reprint,superscriptaddress]{revtex4-1}

\usepackage[T1]{fontenc}
\usepackage{amsmath}
\usepackage{amssymb}
\usepackage{graphicx}
\usepackage{braket}
\usepackage{color}

\newcommand{\op}[1]{\ensuremath{\mathrm{#1}}}
\newcommand{\chitwo}{\ensuremath{\chi^{(2)}} }

\newcommand{\hc}{\textrm{h.~c.}}

\newcommand{\func}[2]{#1\!\left(#2\right)}

\newcommand{\vac}{\ket{0}}
\newcommand{\mapfunc}{\ensuremath{\func{f}{\omega_i, \omega_o}}}
\newcommand{\pump}{\func{\alpha}{\omega_o-\omega_i}}
\newcommand{\phasematch}{\func{\Phi}{\omega_o,\omega_i}}

\begin{document}

\title{A Quantum Pulse Gate based on Spectrally Engineered Sum Frequency Generation}

\author{Andreas Eckstein}
\affiliation{Max Planck Institute for the Science of Light, G\"unther-Scharowsky-Strasse 1, 91054 Erlangen, Germany}
\email[]{andreas.eckstein@mpl.mpg.de}

\author{Benjamin Brecht}
\author{Christine Silberhorn}
\affiliation{Applied Physics, University of Paderborn, Warburgerstrasse 100, 33098 Paderborn, Germany}
\affiliation{Max Planck Institute for the Science of Light, G\"unther-Scharowsky-Strasse 1, 91054 Erlangen, Germany}

\date{\today}

\begin{abstract}
We introduce the concept of a quantum pulse gate (QPG), a method for accessing the intrinsic broadband spectral mode structure of ultrafast quantum states of light. This mode structure can now be harnessed for applications in quantum information processing. We propose an implementation in a PPLN waveguide, based on spectrally engineered sum frequency generation (SFG). It allows us to pick well-defined spectral broadband modes from an ultrafast multi-mode state for interconversion to a broadband mode at another frequency. By pulse-shaping the bright SFG pump beam, different orthogonal broadband modes can be addressed individually and extracted with near unit efficiency.
\end{abstract}


\maketitle
Ultrashort pulses of light play an ever-increasing role in modern quantum information and communications. Today, they already enable high data transmission rates for secure quantum key distribution\cite{Dixon10} and high precision positioning and clock synchronization protocols\cite{Giovannetti01}. In recent years there has been increased interest in a finer control over the rich temporal and spectral structure of quantum light pulses, for applications such as ultrafast probing of the temporal wave function of photon pairs\cite{Peer05}, or efficiently coupling single photons to trapped atoms\cite{Stobinska09}. Given direct access, this structure could also be utilized to encode more information into or extract more quantum information from one pulse of light.

It is possible to decompose any pulse form into any complete set of orthogonal basis functions, or broadband modes\cite{Titulaer66}. Thus it can be considered to be made up of an infinite number of temporally overlapping but independent pulses. While for classical light all basis sets are equivalent, for quantum light there may be one special, intrinsic basis choice\cite{Martinelli2003}. For photon pair states this choice determined by a Schmidt decomposition of their bi-photon spectral amplitude into two correlated basis sets of broadband pulse forms, the Schmidt modes\cite{Law2000}. Heralding one of those photons by detecting the other with a single photon detector (SPD), this correlation results in the preparation of a photon in a mixed state of all Schmidt modes present\cite{Grice01}. But with a SPD sensitive to a certain Schmidt mode, it opens up the possibility to prepare pure single photons in the correlated Schmidt mode. Typically though, SPDs and optical detectors in general exhibit very broad spectral response and are not able to discern between different pulse forms.

To compensate for the detectors' shortcomings, one needs to include a filter operation sensitive to broadband modes. It has already been shown that ordinary spectral filters cannot fulfill this role\cite{Rohde07,Branczyk2010}: They always transmit part of all impinging broadband modes at once, and thus cannot be matched to a single broadband mode. A sufficiently narrow spectral filter can be used to select a monochromatic mode, however this way, high purity heralded quantum states are impossible\cite{Laiho09,Branczyk2010}. Also, most of the original beam's brightness as well as its pulse characteristics are lost.

\begin{figure}
  \centering\includegraphics[width=\linewidth]{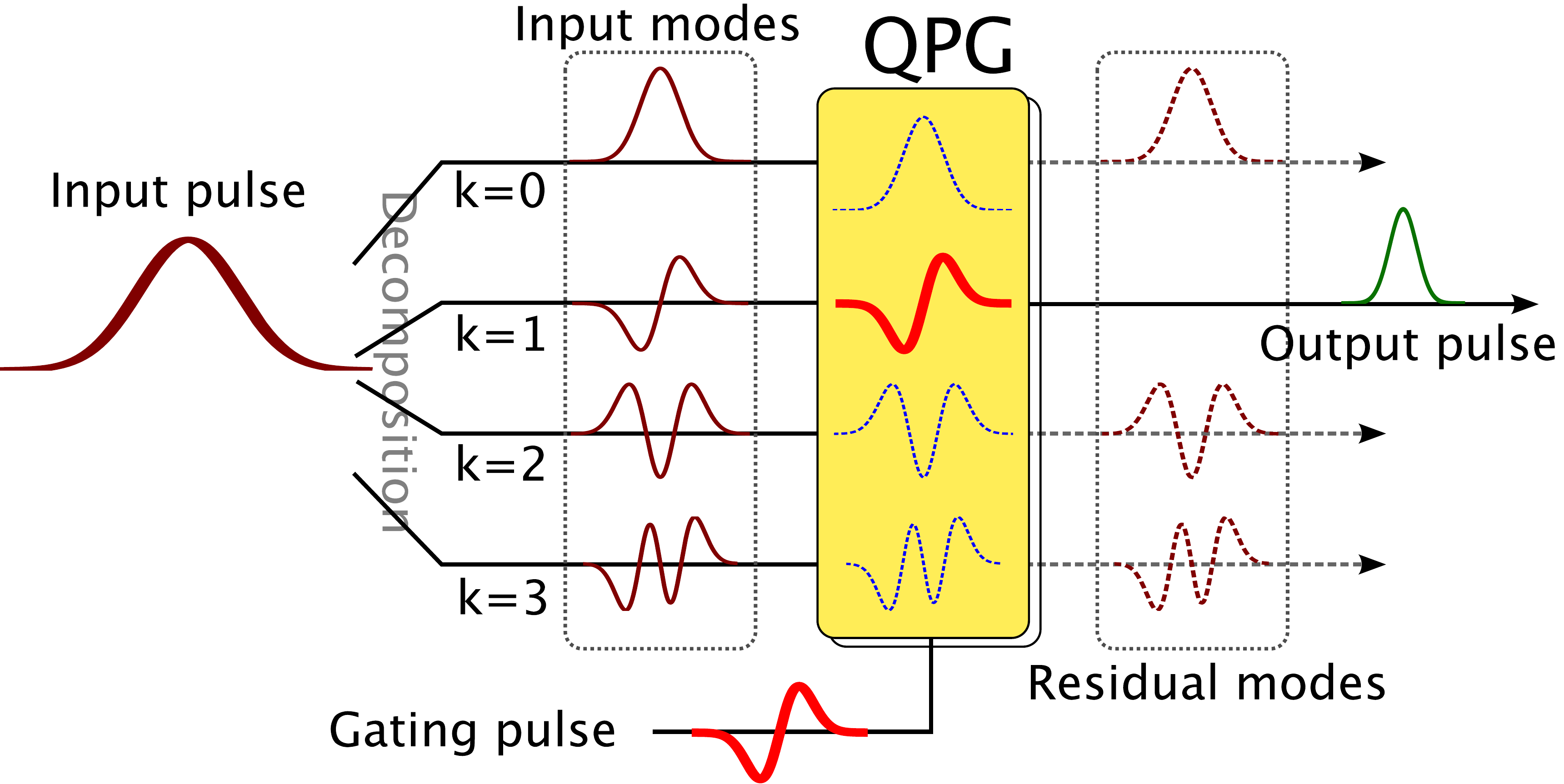}
  \caption{Quantum Pulse Gate schema: Gating with a pulse in spectral broadband mode $u_k$ converts only the corresponding mode from the input pulse to a Gaussian wave packet at sum frequency.}
  \label{fig:schema}
\end{figure}

The idea of using broadband modes as quantum information carriers is especially compelling because of their natural occurrence in ultrafast pulses, and their stability in transmission: Centered around one frequency within a relatively small bandwidth typically, they allow for optical components that are highly optimized for a small spectral range. Since all broadband modes experience the same chromatic dispersion in optical media, they exhibit the same phase modulation and thus stay exactly orthogonal to each other. So a light pulse's broadband mode structure is resilient to the effects of chromatic dispersion, making a multi-channel protocol based on them ideal for optical fiber transmission. Additionally they allow for high transmission rates, as they inherit the ultrashort properties of their 'carrier' pulse, when compared to the 'long' pulses used for classical, narrow-band frequency multiplexing techniques. However, it is extremely challenging to actually access them in a controlled manner: Ordinary spectral filters and standard optical detectors destroy the mode structure of a beam. A homodyne detector with an ultrafast pulsed local oscillator beam is able to select a single broadband mode by spectral overlap, but only at the cost of consuming the whole input beam.\cite{Yuen83,Schumaker84} 

For discrete spatial modes, complete control of a beam's multi-mode structure can be accomplished with linear optics, as combining them to synthesize multi-mode beams and separating constituents without losses is possible\cite{Leach02,*Treps2003,*Lassen2007,*Yarnall07}. In order to exploit the pulse form degree of freedom, we must be able to exact similar control over broadband modes.

An important step towards this goal is to selectively target a single broadband mode for interconversion into a more accessible channel, for instance to shift it to another frequency with SFG. On the single photon level, in the SFG process two single photons ``fuse'' into one photon at their sum frequency inside a $\chi^{\left(2\right)}$-nonlinear material. Well known in classical nonlinear optics, in recent years  it has seen increasing adoption in quantum optics for efficient NIR single photon detection\cite{Roussev04,*Albota04,*Vandevender04,*Tanzilli05}, all-optical fast switching\cite{Vandevender07}, super high resolution timing of quantum pulses\cite{Kuzucu08}, and quantum information erasure\cite{Takesue08}. Moreover, combined with spectral engineering\cite{Grice01,URen2005,*Mosley2008}, it enables a new type of quantum interference between photons of different color\cite{Raymer2010}.

In this Letter we introduce the Quantum Pulse Gate (QPG): A device based on spectrally engineered SFG to extract photons in a well-defined broadband mode from a light beam. We overlap an incoming weak, multi-mode input pulse with a bright, classical gating pulse inside a nonlinear optical material (Fig. \ref{fig:schema}). Spectral engineering ensures that only the fraction of the input pulse which follows the gating pulse form is converted. The residual pulse, orthogonal to the gating pulse, is ignored. An input quantum light pulse's quantum properties can be preserved in conversion by mode-matching the gating pulse to its intrinsic mode structure.

SFG conversion efficiency can be tuned with gating pulse power, and unit efficiency is in principle reachable. Thus the QPG is able to unconditionally filter broadband modes from arbitrary input states, and to convert them into a well-defined Gaussian wave packet at the sum frequency. By pulse-shaping the gating pulse we are able to switch between different target broadband modes during the experiment. By superimposing gating pulses for two different broadband modes, we create interference between those previously orthogonal pulses. In combination with a standard single photon detector we are able to herald pulsed, pure, single-mode single photons from a multi-mode photon pair source.

For a bright classical gating pulse, the effective Hamiltonian of SFG that up-converts a photon in mode '$\op{a}$' to mode '$\op{c}$' is given by
\begin{equation}
	\op{H} = \theta\int\! d\omega_i\ d\omega_o\ \mapfunc\op{a}(\omega_i)
	\op{c}^{\dagger}(\omega_o)+\hc
	\label{eq:effective_hamiltonian}
\end{equation}
Here we introduced the coupling constant $\theta\propto\chi^{(2)} \sqrt{P}$ with $\chi^{(2)}$ denoting the second order nonlinear polarization tensor element of the SFG process and $P$ the gating pulse power. The SFG transfer function $\mapfunc=\pump\times\phasematch$ maps the input frequencies $\omega_i$ to the sum frequencies $\omega_o$, where $\alpha$ is the spectral amplitude of the gating pulse and $\Phi$ the phase matching distribution of the SFG process.

In parametric down-conversion (PDC), the Schmidt decomposition of the joint spectral amplitude of the generated photon pairs reveals their broadband mode structure\cite{Law2000}. Applying the same approach to SFG\cite{Raymer2010} to decompose the spectral transfer function we find
\begin{equation}
	\mapfunc = \sum_k \kappa_k\ \varphi_k(\omega_i)\ \psi_k(\omega_o).
\end{equation}
The decomposition is well-defined and yields two correlated sets of orthonormal spectral amplitude functions $\{\varphi_k(\omega)\}$ and $\{\psi_k(\omega)\}$ and the real Schmidt coefficients $\kappa_k$ which satisfy the relation $\sum_k \kappa_k^2 = 1$. If the gating pulse has the form of a weighted Hermite function $u_k\left(\omega\right)\propto e^{\frac{\left(\omega-\omega_0\right)^2}{2 \sigma^2}} H_k\left(\frac{\omega-\omega_0}{\sigma}\right)$ with $H_k$ the Hermite polynomials, the basis functions of both sets are in good approximation Hermite functions as well. In the Schmidt-decomposed form, the transfer function describes a mapping between pairs of broadband modes $\varphi_k(\omega) \rightarrow \psi_k(\omega)$.

By defining broadband mode operators $\op{A}_k=\int\!d\omega\,\varphi_k\!\left(\omega\right)\,\op{a}\!\left(\omega\right)$ and $\op{C}_k=\int\!d\omega\,\psi_k\!\left(\omega\right)\,\op{c}\!\left(\omega\right)$ corresponding to the Schmidt bases, the effective Hamiltonian from Eq. \ref{eq:effective_hamiltonian} can be rewritten as
\begin{equation}
  \op{H} = \theta \sum_k  \kappa_k \left(\op{A}_k \op{C}_k^\dagger + \op{A}_k^\dagger \op{C}_k \right),
  \label{eq:output_state_schmidt}
\end{equation}

An optical beam splitter has a Hamiltonian of the form $\op{H}_{\textrm{BS}}=\theta\,\op{a}\,\op{c}^\dagger+\hc$ So with respect to broadband modes, SFG can be formally interpreted as a set of beam splitters, independently operating on one pair of broadband modes each, such that $\op{A}_k \rightarrow \cos(\theta_k)\op{A}_k + \imath\sin(\theta_k)\op{C}_k$. The effective coupling constant $\theta_k = \theta\cdot\kappa_k\propto \sqrt{P}$ takes the role of the beam splitter angle. Its transmission probability -- the probability to find a photon in the up-converted mode $\op{C}_k$ if it initially has been in mode $\op{A}_k$ -- is $\eta_k = \sin^2(\theta_k)$.

\begin{figure}
  \includegraphics[width=\linewidth]{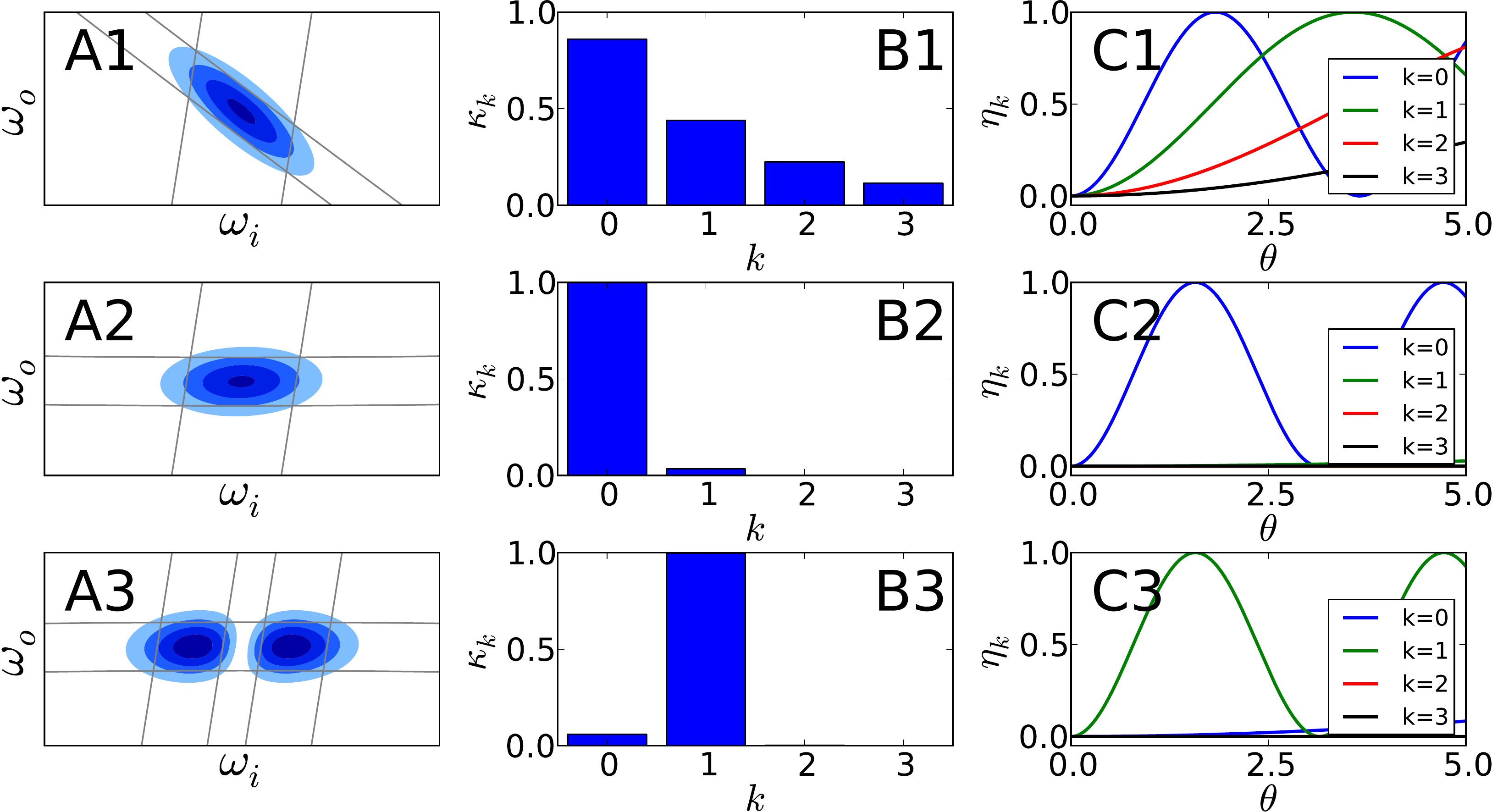}
  \caption{(A1-A3) SFG transfer function \mapfunc\ with (A1) and without (A2,A3) frequency correlations. (B1-B3) Coefficients $\kappa_k$ for the first four Schmidt mode pairs of the transfer functions. (C1-C3) SFG efficiencies $\op{A}_k \rightarrow \op{C}_k$ for the first four Schmidt modes against gating power dependent SFG coupling constant $\theta$}
  \label{fig:jsa_efficiency}
\end{figure}

In Fig. \ref{fig:jsa_efficiency} A1-C1, we illustrate an example for a non-engineered SFG process, as commonly found in pulsed SFG experiments: The transfer function \mapfunc (Fig. \ref{fig:jsa_efficiency} A1) exhibits spectral correlations, causing more than one non-zero Schmidt coefficient (Fig. \ref{fig:jsa_efficiency} B1). This leads to the simultaneous conversion of multiple modes $\op{A}_k$ at once with non-zero coupling constants $\theta_k\propto \sqrt{P}$ for any given gating pulse power $P$ (Fig. \ref{fig:jsa_efficiency} C1). Hence a SFG process in general is not mode-selective.

However, spectral engineering can make SFG mode-selective by eliminating its spectral correlations so that the frequency of an up-converted photon gives no information about its original frequency. Now, Schmidt decomposition yields one predominant parameter $\kappa_k\approx1$ with all others close to zero and a separable transfer function $\mapfunc \approx \kappa_k \varphi_k(\omega_i) \psi_k(\omega_o)$.
Also, now the full coupling $\theta_k\approx\theta$ is exploited, allowing for relatively weak gating beams for unit conversion efficiency. We achieve this by choosing a SFG process with an already correlation-free phasematching function $\Phi$. If the phasematching bandwidth is narrow compared to gating pulse width, spectral correlations are negligible (Fig. \ref{fig:jsa_efficiency} A2-A3), and we can approximate a separable transfer function (Fig. \ref{fig:jsa_efficiency} B2-B3). The effective SFG Hamiltonian is now formally a beam splitter Hamiltonian
\begin{equation}
\op{H}_\textrm{QPG}=\theta_k \op{A}_k \op{C}_0^{\dagger}+\hc\label{qpghamiltonian}
\end{equation}
meaning that only mode $\op{A}_k$ is accepted for conversion. Because of the horizontal phasematching, the target mode is always the Gaussian pulse $\op{C}_0$.

This process implements the QPG, with the bright input pulse used as gate pulse to select a specific broadband mode. By tuning the central wavelength and spectral distribution of the gating pulse, we can control the selected broadband mode's shape, width and central wavelength. We compare the effect of different gating pulse forms: Gating with mode $u_0$ (i.~e. a Gaussian spectrum, Fig. \ref{fig:jsa_efficiency} A2-C2) selects input mode $\op{A}_0$, gating with mode $u_1$ (Fig. \ref{fig:jsa_efficiency} A3-C3) selects $\op{A}_1$ from the input pulse for frequency up-conversion.

\begin{figure}
\includegraphics[width=\linewidth]{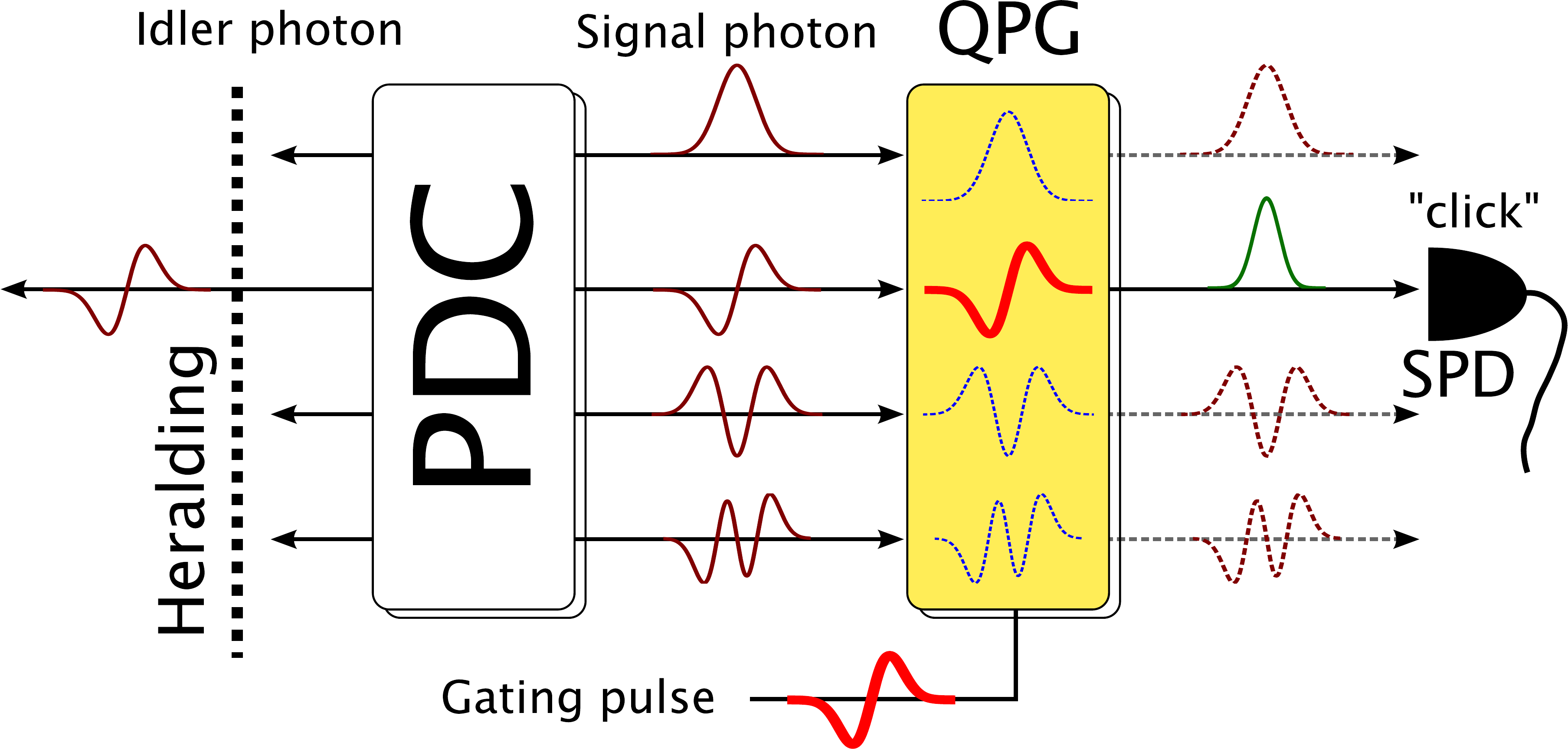}
\caption{A QPG application: Generating pure heralded broadband single photons in different modes from a PDC source of multi-mode photon pairs}
	\label{fig:modeheralding}
\end{figure}

Pure heralded single photons are a crucial resource in many quantum optical applications, but the widely used PDC photon pair sources emit mixed heralded photons in general. We now consider the application of the QPG to ``purify'' those photons. In type-II PDC, a pump photon decays inside a \chitwo-nonlinear medium into one horizontally polarized signal and one vertically polarized idler photon. For a collinear type-II PDC source pumped by ultrafast pulses the general effective Hamiltonian in terms of broadband modes reads
\begin{equation} 
	\op{H}_{\textrm{PDC}}=\chi\sum_k{c_k\left( \op{\tilde{A}}_k^\dagger \op{B}_k^\dagger + \op{\tilde{A}}_k \op{B}_k\right)}.
	\label{eq:hamiltonian_squeezing}
\end{equation}
Using such a photon pair source for the preparation of heralded single photons, one finds that those are usually not in pure, but spectrally mixed states\cite{Grice01}, and thus of limited usefulness for most quantum optical applications. We feed the signal photon (containing all broadband modes $\op{\tilde{A}}_k$) from the PDC source into the QPG which is mode-matched such that $\op{\tilde{A}}_0=\op{A}_0$. We note that for heralding pure single photons or pure Fock states\cite{Rohde07}, mode-matching is not necessary and an engineered SFG process according to Eq. \ref{qpghamiltonian} is sufficient. In that case however, the resulting pulse shape is a coherent superposition of all input modes. Here, only the 0th mode is selected, and the higher modes do not interact with the QPG because the according beam splitter transformations yield the identity $\op{A}_k\rightarrow \op{A}_k$ for $k>0$. We choose the gating pulse power such that $\theta_0=\frac{\pi}{2}$ for optimal conversion efficiency. Combining the PDC source with a subsequent QPG results transforms the PDC Hamiltonian as $\op{H}_{\textrm{PDC}}\rightarrow \op{H}^\prime= e^{-\imath \op{H}_\textrm{QPG}}\op{H}_{\textrm{PDC}}e^{\imath \op{H}_\textrm{QPG}}$, and we obtain
\begin{equation}
    \op{H}^\prime= \imath\chi \op{B}_0^\dagger \op{C}_0^\dagger+\chi\sum_{k=1}^\infty c_k \op{\tilde{A}}_k^\dagger \op{B}_k^\dagger +\hc
    \label{eq:upstate}
\end{equation}
Since mode $\op{C}_0$ is centered at the sum frequency of input and gating pulse, it can be split off easily into a separate beam path with a dichroic mirror. Conditioning on single photon events on the path of $\op{C}_0$ provides us with pure heralded single photons in mode $\op{B}_0$. Fig. \ref{fig:modeheralding} illustrates this scheme: A photon detection event heralds a pure single photon pulse in broadband mode $u_1$. This process can be cascaded to successively pick off several modes $\op{A}_k$ from the input beam. Note that if we insert a mode matched QPG into the vertically polarized PDC beam to convert $\op{B}_0$ into $\op{D}_0$, we can unconditionally single out an ultrafast two-mode squeezed vacuum state $e^{\imath \chi c_0 \left(\op{C}_0^\dagger \op{D}_0^\dagger+\op{C}_0 \op{D}_0\right)}\vac$ from a multi-mode squeezer\cite{Eckstein11}.

Finally we give realistic parameters to show the feasibility of an experimental implementation of the QPG, and analyze the mode selection  performance. For the SFG process we use a periodically poled LiNbO$_3$ (PPLN) waveguide with an area of $8 \times 5\mu\mathrm{m}^2$, a length of L=50mm, a $\Lambda=4.2\mu\mathrm{m}$ periodic poling period and at $175^{\circ}$C to achieve phasematching for SFG of an input pulse at 1550nm to about 557nm. It is gated by coherent laser pulses at 870nm with 2ps pulse length or a spectrum with 0.635nm FWHM to ensure a transfer function separability. The uncorrelated, separable transfer functions in Fig. \ref{fig:jsa_efficiency} (A2-A3) are calculated from these parameters, using gating pulses with $u_0$ and $u_1$ as spectral amplitude, respectively. 
\begin{figure}
\includegraphics[width=\linewidth]{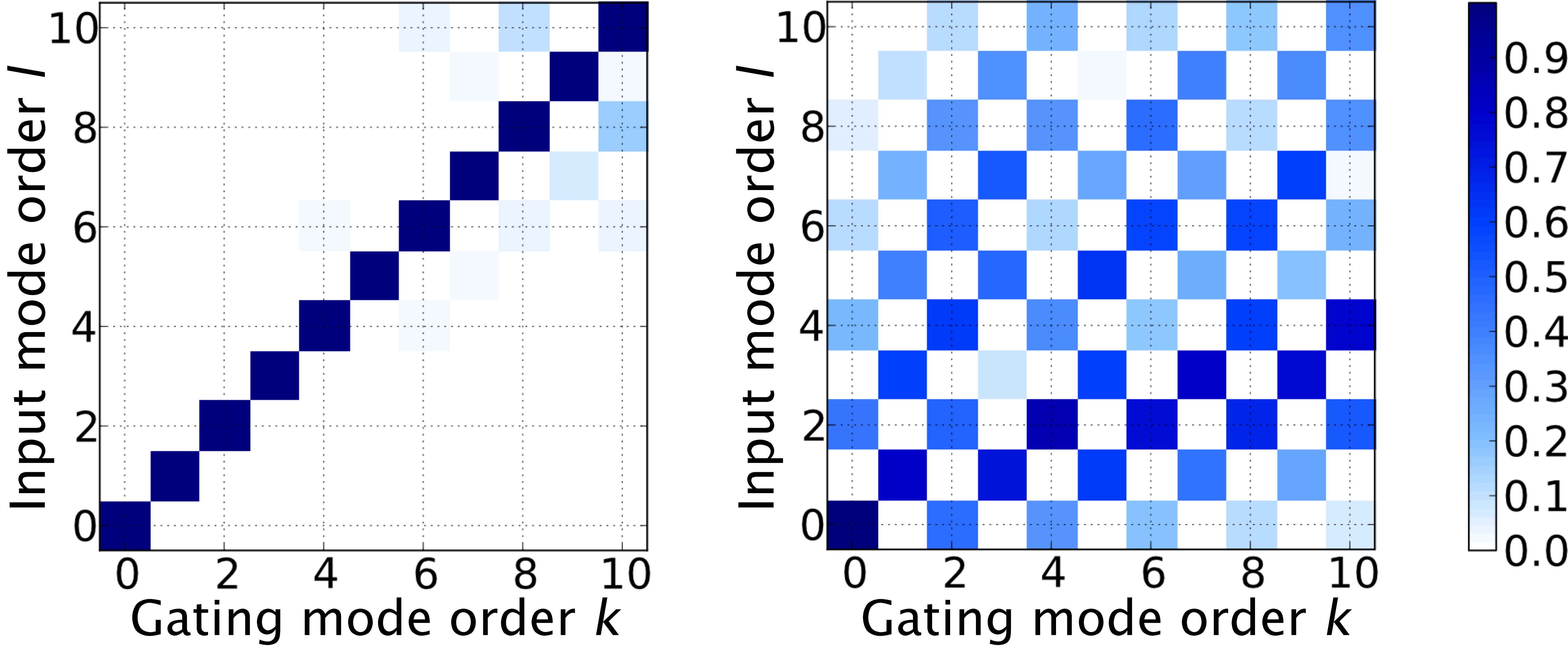}
\caption{Overlap between input pulse mode $\tilde{u}_l$ and QPG Schmidt mode $\varphi_k$ for mode-matched (left) and non-mode-matched (right) case.}
	\label{fig:cross}
\end{figure}

In Fig. \ref{fig:cross} we illustrate the switching capabilities of our QPG, as well as the impact of mode matching. For the given material parameters, we employ gating pulses with pulse form $u_0$ to $u_{10}$, determine the Schmidt decomposition of the resulting transfer function \mapfunc, and plot the overlap of the predominant Schmidt function $\varphi_k$ ($\kappa_k\approx 1$) with an Hermitian input mode $\tilde{u}_l$ from an incident light pulse. On the left, gating and input pulse have equal frequency FWHM, which is essential for good mode matching. Now, by switching the order $k$ of the gating mode (and without changing the physical parameters of the QPG), we select with high fidelity only the input mode $k$ to be converted. For $k\leq10$, the overlap $\left|\int\!d\omega \tilde{u}_k^*\!\left(\omega\right)\varphi_k\!\left(\omega\right)\right|^2$ exceeds 99\%, and the overlap for all other input modes combined therefore is less than 1\%: Only a negligible fraction of modes other than the selected input mode are converted.

In contrast, Fig. \ref{fig:cross} (right) has no mode matching, the gating pulse FWHM is twice that of the input pulse. Multiple strong overlaps between SFG Schmidt modes $\varphi_k$ and input modes $\tilde{u}_l$ appear: A wide range of modes is converted for any given input spectrum. The checkerboard pattern reflects the fact that only Hermite modes of the same parity overlap, and the SFG Schmidt modes are in good approximation Hermite modes.

In conclusion, we have introduced the concept of the QPG, a flexible device to split well-defined broadband modes from a light pulse based on spectrally engineered SFG. The selected mode can be switched by shaping the gating pulse spectrum and converted with high efficiency. Further, we have given a realistic set of experimental parameters for a QPG realized in a PPLN waveguide and demonstrated the high flexibility of the QPG achieved through shaping the gating pulse form. We proposed as an initial application the preparation of pure heralded single photons from an arbitrary type II PDC source. For pulsed QKD schemes\cite{Dixon10}, it can act as a de-multiplexer of multiple quantum channels within one physical pulse, and in metrology it may be used to further enhance measurement accuracy beyond the classical limit by replacing multiple squeezed pulses\cite{Giovannetti01} with one multi-mode squeezed pulse of light.

We would like to thank Andreas Christ$^2$ for helpful discussions and acknowledge support of this work under the EC grant agreement CORNER (FP7-ICT-213681).


\begin{thebibliography}{10}%
\makeatletter
\providecommand \@ifxundefined [1]{%
 \ifx #1\undefined \expandafter \@firstoftwo
 \else \expandafter \@secondoftwo
\fi
}%
\providecommand \@ifnum [1]{%
 \ifnum #1\expandafter \@firstoftwo
 \else \expandafter \@secondoftwo
\fi
}%
\providecommand \enquote [1]{``#1''}%
\providecommand \bibnamefont  [1]{#1}%
\providecommand \bibfnamefont [1]{#1}%
\providecommand \citenamefont [1]{#1}%
\providecommand\href[0]{\@sanitize\@href}%
\providecommand\@href[1]{\endgroup\@@startlink{#1}\endgroup\@@href}%
\providecommand\@@href[1]{#1\@@endlink}%
\providecommand \@sanitize [0]{\begingroup\catcode`\&12\catcode`\#12\relax}%
\@ifxundefined \pdfoutput {\@firstoftwo}{%
 \@ifnum{\z@=\pdfoutput}{\@firstoftwo}{\@secondoftwo}%
}{%
 \providecommand\@@startlink[1]{\leavevmode}%
 \providecommand\@@endlink[0]{}%
}{%
 \providecommand\@@startlink[1]{%
  \leavevmode
  \pdfstartlink
   attr{/Border[0 0 1 ]/H/I/C[0 1 1]}%
   user{/Subtype/Link/A<</Type/Action/S/URI/URI(#1)>>}%
  \relax
 }%
 \providecommand\@@endlink[0]{\pdfendlink}%
}%
\providecommand \url  [0]{\begingroup\@sanitize \@url }%
\providecommand \@url [1]{\endgroup\@href {#1}{\urlprefix}}%
\providecommand \urlprefix [0]{URL }%
\providecommand \Eprint[0]{\href }%
\@ifxundefined \urlstyle {%
  \providecommand \doi [1]{doi:\discretionary{}{}{}#1}%
}{%
  \providecommand \doi [0]{doi:\discretionary{}{}{}\begingroup
  \urlstyle{rm}\Url }%
}%
\providecommand \doibase [0]{http://dx.doi.org/}%
\providecommand \Doi[1]{\href{\doibase#1}}%
\providecommand \bibAnnote [3]{%
  \BibitemShut{#1}%
  \begin{quotation}\noindent
    \textsc{Key:}\ #2\\\textsc{Annotation:}\ #3%
  \end{quotation}%
}%
\providecommand \bibAnnoteFile [2]{%
  \IfFileExists{#2}{\bibAnnote {#1} {#2} {\input{#2}}}{}%
}%
\providecommand \typeout [0]{\immediate \write \m@ne }%
\providecommand \selectlanguage [0]{\@gobble}%
\providecommand \bibinfo [0]{\@secondoftwo}%
\providecommand \bibfield [0]{\@secondoftwo}%
\providecommand \translation [1]{[#1]}%
\providecommand \BibitemOpen[0]{}%
\providecommand \bibitemStop [0]{}%
\providecommand \bibitemNoStop [0]{.\EOS\space}%
\providecommand \EOS [0]{\spacefactor3000\relax}%
\providecommand \BibitemShut [1]{\csname bibitem#1\endcsname}%
\bibitem{Dixon10}%
  \BibitemOpen
  \bibfield{author}{%
  \bibinfo {author} {\bibfnamefont{A.~R.}\ \bibnamefont{Dixon}} \emph{et~al.},\
  }%
  \bibfield{journal}{%
  \Doi{10.1063/1.3385293}{\bibinfo {journal} {Applied Physics Letters}}\ }%
  \textbf{\bibinfo {volume} {96}},\ \bibinfo {eid} {161102} (\bibinfo {year}
  {2010})%
  \bibAnnoteFile{NoStop}{Dixon10}%
\bibitem{Giovannetti01}%
  \BibitemOpen
  \bibfield{author}{%
  \bibinfo {author} {\bibfnamefont{V.}~\bibnamefont{Giovannetti}}, \bibinfo
  {author} {\bibfnamefont{S.}~\bibnamefont{Lloyd}},\ and\ \bibinfo {author}
  {\bibfnamefont{L.}~\bibnamefont{Maccone}},\ }%
  \bibfield{journal}{%
  \Doi{10.1038/35086525}{\bibinfo {journal} {Nature}}\ }%
  \textbf{\bibinfo {volume} {412}},\ \bibinfo {pages} {417} (\bibinfo {month}
  {Jul.}\ \bibinfo {year} {2001}),\ ISSN \bibinfo {issn} {0028-0836}%
  \bibAnnoteFile{NoStop}{Giovannetti01}%
\bibitem{Peer05}%
  \BibitemOpen
  \bibfield{author}{%
  \bibinfo {author} {\bibfnamefont{A.}~\bibnamefont{Pe'er}}, \bibinfo {author}
  {\bibfnamefont{B.}~\bibnamefont{Dayan}}, \bibinfo {author}
  {\bibfnamefont{A.~A.}\ \bibnamefont{Friesem}},\ and\ \bibinfo {author}
  {\bibfnamefont{Y.}~\bibnamefont{Silberberg}},\ }%
  \bibfield{journal}{%
  \Doi{10.1103/PhysRevLett.94.073601}{\bibinfo {journal} {Phys. Rev. Lett.}}\
  }%
  \textbf{\bibinfo {volume} {94}},\ \bibinfo {pages} {073601} (\bibinfo {month}
  {Feb}\ \bibinfo {year} {2005})%
  \bibAnnoteFile{NoStop}{Peer05}%
\bibitem{Stobinska09}%
  \BibitemOpen
  \bibfield{author}{%
  \bibinfo {author} {\bibfnamefont{M.}~\bibnamefont{Stobi\'nska}}, \bibinfo
  {author} {\bibfnamefont{G.}~\bibnamefont{Alber}},\ and\ \bibinfo {author}
  {\bibfnamefont{G.}~\bibnamefont{Leuchs}},\ }%
  \bibfield{journal}{%
  \Doi{10.1209/0295-5075/86/14007}{\bibinfo {journal} {Europhys. Lett.}}\ }%
  \textbf{\bibinfo {volume} {86}},\ \bibinfo {pages} {14007} (\bibinfo {year}
  {2009})%
  \bibAnnoteFile{NoStop}{Stobinska09}%
\bibitem{Titulaer66}%
  \BibitemOpen
  \bibfield{author}{%
  \bibinfo {author} {\bibfnamefont{U.~M.}\ \bibnamefont{Titulaer}}\ and\
  \bibinfo {author} {\bibfnamefont{R.~J.}\ \bibnamefont{Glauber}},\ }%
  \bibfield{journal}{%
  \Doi{10.1103/PhysRev.145.1041}{\bibinfo {journal} {Phys. Rev.}}\ }%
  \textbf{\bibinfo {volume} {145}},\ \bibinfo {pages} {1041} (\bibinfo {month}
  {May}\ \bibinfo {year} {1966})%
  \bibAnnoteFile{NoStop}{Titulaer66}%
\bibitem{Martinelli2003}%
  \BibitemOpen
  \bibfield{author}{%
  \bibinfo {author} {\bibfnamefont{M.}~\bibnamefont{Martinelli}}
  \emph{et~al.},\ }%
  \bibfield{journal}{%
  \Doi{10.1103/PhysRevA.67.023808}{\bibinfo {journal} {Phys. Rev. A}}\ }%
  \textbf{\bibinfo {volume} {67}},\ \bibinfo {pages} {023808} (\bibinfo {year}
  {2003})%
  \bibAnnoteFile{NoStop}{Martinelli2003}%
\bibitem{Law2000}%
  \BibitemOpen
  \bibfield{author}{%
  \bibinfo {author} {\bibfnamefont{C.~K.}\ \bibnamefont{Law}}, \bibinfo
  {author} {\bibfnamefont{I.~A.}\ \bibnamefont{Walmsley}},\ and\ \bibinfo
  {author} {\bibfnamefont{J.~H.}\ \bibnamefont{Eberly}},\ }%
  \bibfield{journal}{%
  \Doi{10.1103/PhysRevLett.84.5304}{\bibinfo {journal} {Phys. Rev. Lett.}}\ }%
  \textbf{\bibinfo {volume} {84}},\ \bibinfo {pages} {5304} (\bibinfo {year}
  {2000})%
  \bibAnnoteFile{NoStop}{Law2000}%
\bibitem{Grice01}%
  \BibitemOpen
  \bibfield{author}{%
  \bibinfo {author} {\bibfnamefont{W.~P.}\ \bibnamefont{Grice}}, \bibinfo
  {author} {\bibfnamefont{A.~B.}\ \bibnamefont{U'Ren}},\ and\ \bibinfo {author}
  {\bibfnamefont{I.~A.}\ \bibnamefont{Walmsley}},\ }%
  \bibfield{journal}{%
  \Doi{10.1103/PhysRevA.64.063815}{\bibinfo {journal} {Phys. Rev. A}}\ }%
  \textbf{\bibinfo {volume} {64}},\ \bibinfo {pages} {063815} (\bibinfo {month}
  {Nov}\ \bibinfo {year} {2001})%
  \bibAnnoteFile{NoStop}{Grice01}%
\bibitem{Rohde07}%
  \BibitemOpen
  \bibfield{author}{%
  \bibinfo {author} {\bibfnamefont{P.~P.}\ \bibnamefont{Rohde}}, \bibinfo
  {author} {\bibfnamefont{W.}~\bibnamefont{Mauerer}},\ and\ \bibinfo {author}
  {\bibfnamefont{C.}~\bibnamefont{Silberhorn}},\ }%
  \bibfield{journal}{%
  \Doi{10.1088/1367-2630/9/4/091}{\bibinfo {journal} {New Journal of Physics}}\
  }%
  \textbf{\bibinfo {volume} {9}},\ \bibinfo {pages} {91} (\bibinfo {year}
  {2007})%
  \bibAnnoteFile{NoStop}{Rohde07}%
\bibitem{Branczyk2010}%
  \BibitemOpen
  \bibfield{author}{%
  \bibinfo {author} {\bibfnamefont{A.~M.}\ \bibnamefont{Branczyk}}, \bibinfo
  {author} {\bibfnamefont{T.~C.}\ \bibnamefont{Ralph}}, \bibinfo {author}
  {\bibfnamefont{W.}~\bibnamefont{Helwig}},\ and\ \bibinfo {author}
  {\bibfnamefont{C.}~\bibnamefont{Silberhorn}},\ }%
  \bibfield{journal}{%
  \Doi{10.1088/1367-2630/12/6/063001}{\bibinfo {journal} {New Journal of
  Physics}}\ }%
  \textbf{\bibinfo {volume} {12}},\ \bibinfo {pages} {063001} (\bibinfo {year}
  {2010})%
  \bibAnnoteFile{NoStop}{Branczyk2010}%
\bibitem{Laiho09}%
  \BibitemOpen
  \bibfield{author}{%
  \bibinfo {author} {\bibfnamefont{K.}~\bibnamefont{Laiho}}, \bibinfo {author}
  {\bibfnamefont{K.~N.}\ \bibnamefont{Cassemiro}},\ and\ \bibinfo {author}
  {\bibfnamefont{C.}~\bibnamefont{Silberhorn}},\ }%
  \bibfield{journal}{%
  \Doi{10.1364/OE.17.022823}{\bibinfo {journal} {Opt. Express}}\ }%
  \textbf{\bibinfo {volume} {17}},\ \bibinfo {pages} {22823} (\bibinfo {month}
  {Dec}\ \bibinfo {year} {2009})%
  \bibAnnoteFile{NoStop}{Laiho09}%
\bibitem{Yuen83}%
  \BibitemOpen
  \bibfield{author}{%
  \bibinfo {author} {\bibfnamefont{H.~P.}\ \bibnamefont{Yuen}}\ and\ \bibinfo
  {author} {\bibfnamefont{V.~W.~S.}\ \bibnamefont{Chan}},\ }%
  \bibfield{journal}{%
  \Doi{10.1364/OL.8.000177}{\bibinfo {journal} {Opt. Lett.}}\ }%
  \textbf{\bibinfo {volume} {8}},\ \bibinfo {pages} {177} (\bibinfo {year}
  {1983})%
  \bibAnnoteFile{NoStop}{Yuen83}%
\bibitem{Schumaker84}%
  \BibitemOpen
  \bibfield{author}{%
  \bibinfo {author} {\bibfnamefont{B.~L.}\ \bibnamefont{Schumaker}},\ }%
  \bibfield{journal}{%
  \Doi{10.1364/OL.9.000189}{\bibinfo {journal} {Opt. Lett.}}\ }%
  \textbf{\bibinfo {volume} {9}},\ \bibinfo {pages} {189} (\bibinfo {year}
  {1984})%
  \bibAnnoteFile{NoStop}{Schumaker84}%
\bibitem{Leach02}%
  \BibitemOpen
  \bibfield{author}{%
  \bibinfo {author} {\bibfnamefont{J.}~\bibnamefont{Leach}} \emph{et~al.},\ }%
  \bibfield{journal}{%
  \Doi{10.1103/PhysRevLett.88.257901}{\bibinfo {journal} {Phys. Rev. Lett.}}\
  }%
  \textbf{\bibinfo {volume} {88}},\ \bibinfo {pages} {257901} (\bibinfo {month}
  {Jun}\ \bibinfo {year} {2002})%
  \bibAnnoteFile{NoStop}{Leach02}%
\bibitem{Treps2003}%
  \BibitemOpen
  \bibfield{author}{%
  \bibinfo {author} {\bibfnamefont{N.}~\bibnamefont{Treps}} \emph{et~al.},\ }%
  \bibfield{journal}{%
  \Doi{10.1126/science.1086489}{\bibinfo {journal} {Science}}\ }%
  \textbf{\bibinfo {volume} {301}},\ \bibinfo {pages} {940} (\bibinfo {year}
  {2003})%
  \bibAnnoteFile{NoStop}{Treps2003}%
\bibitem{Lassen2007}%
  \BibitemOpen
  \bibfield{author}{%
  \bibinfo {author} {\bibfnamefont{M.}~\bibnamefont{Lassen}} \emph{et~al.},\ }%
  \bibfield{journal}{%
  \Doi{10.1103/PhysRevLett.98.083602}{\bibinfo {journal} {Phys. Rev. Lett.}}\
  }%
  \textbf{\bibinfo {volume} {98}},\ \bibinfo {pages} {083602} (\bibinfo {year}
  {2007})%
  \bibAnnoteFile{NoStop}{Lassen2007}%
\bibitem{Yarnall07}%
  \BibitemOpen
  \bibfield{author}{%
  \bibinfo {author} {\bibfnamefont{T.}~\bibnamefont{Yarnall}}, \bibinfo
  {author} {\bibfnamefont{A.~F.}\ \bibnamefont{Abouraddy}}, \bibinfo {author}
  {\bibfnamefont{B.~E.~A.}\ \bibnamefont{Saleh}},\ and\ \bibinfo {author}
  {\bibfnamefont{M.~C.}\ \bibnamefont{Teich}},\ }%
  \bibfield{journal}{%
  \Doi{10.1103/PhysRevLett.99.250502}{\bibinfo {journal} {Phys. Rev. Lett.}}\
  }%
  \textbf{\bibinfo {volume} {99}},\ \bibinfo {pages} {250502} (\bibinfo {month}
  {Dec}\ \bibinfo {year} {2007})%
  \bibAnnoteFile{NoStop}{Yarnall07}%
\bibitem{Roussev04}%
  \BibitemOpen
  \bibfield{author}{%
  \bibinfo {author} {\bibfnamefont{R.~V.}\ \bibnamefont{Roussev}}, \bibinfo
  {author} {\bibfnamefont{C.}~\bibnamefont{Langrock}}, \bibinfo {author}
  {\bibfnamefont{J.~R.}\ \bibnamefont{Kurz}},\ and\ \bibinfo {author}
  {\bibfnamefont{M.~M.}\ \bibnamefont{Fejer}},\ }%
  \bibfield{journal}{%
  \Doi{10.1364/OL.29.001518}{\bibinfo {journal} {Opt. Lett.}}\ }%
  \textbf{\bibinfo {volume} {29}},\ \bibinfo {pages} {1518} (\bibinfo {year}
  {2004})%
  \bibAnnoteFile{NoStop}{Roussev04}%
\bibitem{Albota04}%
  \BibitemOpen
  \bibfield{author}{%
  \bibinfo {author} {\bibfnamefont{M.~A.}\ \bibnamefont{Albota}}\ and\ \bibinfo
  {author} {\bibfnamefont{F.~C.}\ \bibnamefont{Wong}},\ }%
  \bibfield{journal}{%
  \Doi{10.1364/OL.29.001449}{\bibinfo {journal} {Opt. Lett.}}\ }%
  \textbf{\bibinfo {volume} {29}},\ \bibinfo {pages} {1449} (\bibinfo {year}
  {2004})%
  \bibAnnoteFile{NoStop}{Albota04}%
\bibitem{Vandevender04}%
  \BibitemOpen
  \bibfield{author}{%
  \bibinfo {author} {\bibfnamefont{A.~P.}\ \bibnamefont{Vandevender}}\ and\
  \bibinfo {author} {\bibfnamefont{P.~G.}\ \bibnamefont{Kwiat}},\ }%
  \bibfield{journal}{%
  \Doi{10.1080/09500340408235283}{\bibinfo {journal} {J. Mod. Opt.}}\ }%
  \textbf{\bibinfo {volume} {51}},\ \bibinfo {pages} {1433} (\bibinfo {year}
  {2004}),\ ISSN \bibinfo {issn} {0950-0340}%
  \bibAnnoteFile{NoStop}{Vandevender04}%
\bibitem{Tanzilli05}%
  \BibitemOpen
  \bibfield{author}{%
  \bibinfo {author} {\bibfnamefont{S.}~\bibnamefont{Tanzilli}} \emph{et~al.},\
  }%
  \bibfield{journal}{%
  \Doi{10.1038/nature04009}{\bibinfo {journal} {Nature}}\ }%
  \textbf{\bibinfo {volume} {437}},\ \bibinfo {pages} {116} (\bibinfo {year}
  {2005}),\ ISSN \bibinfo {issn} {0028-0836}%
  \bibAnnoteFile{NoStop}{Tanzilli05}%
\bibitem{Vandevender07}%
  \BibitemOpen
  \bibfield{author}{%
  \bibinfo {author} {\bibfnamefont{A.~P.}\ \bibnamefont{VanDevender}}\ and\
  \bibinfo {author} {\bibfnamefont{P.~G.}\ \bibnamefont{Kwiat}},\ }%
  \bibfield{journal}{%
  \Doi{10.1364/OE.15.004677}{\bibinfo {journal} {Opt. Express}}\ }%
  \textbf{\bibinfo {volume} {15}},\ \bibinfo {pages} {4677} (\bibinfo {year}
  {2007})%
  \bibAnnoteFile{NoStop}{Vandevender07}%
\bibitem{Kuzucu08}%
  \BibitemOpen
  \bibfield{author}{%
  \bibinfo {author} {\bibfnamefont{O.}~\bibnamefont{Kuzucu}}, \bibinfo {author}
  {\bibfnamefont{F.~N.~C.}\ \bibnamefont{Wong}}, \bibinfo {author}
  {\bibfnamefont{S.}~\bibnamefont{Kurimura}},\ and\ \bibinfo {author}
  {\bibfnamefont{S.}~\bibnamefont{Tovstonog}},\ }%
  \bibfield{journal}{%
  \Doi{10.1103/PhysRevLett.101.153602}{\bibinfo {journal} {Phys. Rev. Lett.}}\
  }%
  \textbf{\bibinfo {volume} {101}},\ \bibinfo {pages} {153602} (\bibinfo {year}
  {2008})%
  \bibAnnoteFile{NoStop}{Kuzucu08}%
\bibitem{Takesue08}%
  \BibitemOpen
  \bibfield{author}{%
  \bibinfo {author} {\bibfnamefont{H.}~\bibnamefont{Takesue}},\ }%
  \bibfield{journal}{%
  \Doi{10.1103/PhysRevLett.101.173901}{\bibinfo {journal} {Phys. Rev. Lett.}}\
  }%
  \textbf{\bibinfo {volume} {101}},\ \bibinfo {pages} {173901} (\bibinfo
  {month} {Oct}\ \bibinfo {year} {2008})%
  \bibAnnoteFile{NoStop}{Takesue08}%
\bibitem{URen2005}%
  \BibitemOpen
  \bibfield{author}{%
  \bibinfo {author} {\bibfnamefont{A.~B.}\ \bibnamefont{U'Ren}} \emph{et~al.},\
  }%
  \bibfield{journal}{%
  \bibinfo {journal} {Laser Physics}\ }%
  \textbf{\bibinfo {volume} {15}} (\bibinfo {year} {2005}),\ ISSN \bibinfo
  {issn} {{1054-660X}}%
  \bibAnnoteFile{NoStop}{URen2005}%
\bibitem{Mosley2008}%
  \BibitemOpen
  \bibfield{author}{%
  \bibinfo {author} {\bibfnamefont{P.~J.}\ \bibnamefont{Mosley}}
  \emph{et~al.},\ }%
  \bibfield{journal}{%
  \Doi{10.1103/PhysRevLett.100.133601}{\bibinfo {journal} {Phys. Rev. Lett.}}\
  }%
  \textbf{\bibinfo {volume} {100}},\ \bibinfo {eid} {133601} (\bibinfo {year}
  {2008})%
  \bibAnnoteFile{NoStop}{Mosley2008}%
\bibitem{Raymer2010}%
  \BibitemOpen
  \bibfield{author}{%
  \bibinfo {author} {\bibfnamefont{M.~G.}\ \bibnamefont{Raymer}}, \bibinfo
  {author} {\bibfnamefont{S.~J.}\ \bibnamefont{van Enk}}, \bibinfo {author}
  {\bibfnamefont{C.~J.}\ \bibnamefont{McKinstrie}},\ and\ \bibinfo {author}
  {\bibfnamefont{H.~J.}\ \bibnamefont{McGuinness}},\ }%
  \bibfield{journal}{%
  \Doi{10.1016/j.optcom.2009.10.057}{\bibinfo {journal} {Opt. Comm.}}\ }%
  \textbf{\bibinfo {volume} {283}},\ \bibinfo {pages} {747 } (\bibinfo {year}
  {2010})%
  \bibAnnoteFile{NoStop}{Raymer2010}%
\bibitem{Eckstein11}%
  \BibitemOpen
  \bibfield{author}{%
  \bibinfo {author} {\bibfnamefont{A.}~\bibnamefont{Eckstein}}, \bibinfo
  {author} {\bibfnamefont{A.}~\bibnamefont{Christ}}, \bibinfo {author}
  {\bibfnamefont{P.~J.}\ \bibnamefont{Mosley}},\ and\ \bibinfo {author}
  {\bibfnamefont{C.}~\bibnamefont{Silberhorn}},\ }%
  \bibfield{journal}{%
  \Doi{10.1103/PhysRevLett.106.013603}{\bibinfo {journal} {Phys. Rev. Lett.}}\
  }%
  \textbf{\bibinfo {volume} {106}},\ \bibinfo {pages} {013603} (\bibinfo
  {month} {Jan}\ \bibinfo {year} {2011})%
  \bibAnnoteFile{NoStop}{Eckstein11}%
\end{thebibliography}

%

\end{document}